\documentclass[12pt,preprint]{aastex} 

\shorttitle{Dust Coagulation in a Non-Uniform Nebula}
\shortauthors{Haghighipour}

\begin{document}

\title{Rapid Growth of Dust Particles in the Vicinity of
Local Gas Density Enhancements in a Solar Nebula}

\author{Nader Haghighipour}
\affil{Department of Terrestrial Magnetism and NASA Astrobiology Institute, \\
Carnegie Institution of Washington, 5241 Broad Branch Road, \\
Washington, DC 20015}

\email{nader@dtm.ciw.edu}

\begin{abstract}

The results of a numerical study of the growth
of solid particles, ranging from 1 micron to 1 millimeter
in size, in the vicinity of an azimuthally
symmetric density enhancement of a protostellar disk 
are presented. It is shown that the
combined effect of gas drag and pressure gradients,
which causes solid objects to rapidly migrate
toward the location of the local maximum density, 
can also enhance the rate of growth of
dust particles to larger objects. The results of
numerical simulations of such growth processes are presented
and the effects of the changes of the physical properties
of particles on the rates of their growth
are also discussed.
\end{abstract}


\section{Introduction}

The focus of this paper is on the study of
the growth rate of dust grains
in the vicinity of a local maximum density
of a nebula.
It has recently been shown that in a nebula 
with local gas density enhancements,
the combined effect of gas drag and pressure gradients
causes solid objects to rapidly migrate toward the location
of maximum density (Haghighipour $\&$ Boss, 2003a$\&$b,
hereafter HB03a$\&$b). In those papers,
the authors have
studied the dynamics of a single object in a 
non-uniform gaseous disk with
no particulate background material.
The only interaction considered in those papers
between the object and the disk was through the gas drag.
If the disk contains a mixture of gas and
small solid particles, however, the motion of an object
will also be affected by its interaction with this
particulate material. For instance,
an object may sweep up smaller particles and grow
larger during its gas-drag induced migration. 
In this paper, I consider particulate material coupled
to a gaseous disk with a local density enhancement,
and study the rate of growth of 
small particles while sweeping up the particles
of the background material. 

For the past four decades, the growth of dust particles
in a solar nebula was studied by many authors.
In 1969, Safronov studied the growth of dust particles
during their vertical descent in a quiescent nebula and
showed that, considering their sticking growth, settling of
dust particles could be quite rapid. Safronov's formulation
indicated that dust particles could grow to 1 centimeter in size
at a 1 AU radial distance from the Sun, assuming
a 100$\%$ sticking probability \citep{Safronov}. 
Growth and sedimentation of dust particles have also been studied
extensively by \citet{Weid80,Weid84} and \citet[][1986]{Nakagawa81}.
\citet{Weid80,Weid84} considered coagulation of dust grains due to
van der Waals force in their settling toward the midplane 
and showed that, while such particles grow to 1 centimeter in size,
they create a turbulent layer in the vicinity of the midplane which
increases the rates of their collision and their coalescence to
larger objects. \citet{Nakagawa81} studied the
growth and settling of dust grains at the location of Earth
in a nebula and showed that, similar to the results presented
by Safronov, the settling of particles and the formation
of a layer of centimeter-sized objects in the vicinity of the midplane can 
occur in a relatively short time ($\simeq$300 years).
This study was later extended by \citet{Nakagawa86}
to the regions of Jupiter and Neptune in which the 
authors studied the settling and growth of dust
particles in laminar phase of a nebula and presented
analytical expressions for the rates of the growth and
also vertical and radial motions of solids. By treating
a dust layer as a two-dimensional fluid, \citet{Nakagawa86}
showed that the total settling time of dust grains in
the two stages of gas- and dust-dominant phases are smaller
than those estimated by \citet{Weid80} and \citet{Nakagawa81}.
More recently, \citet{Sup00} have studied the coagulation of
small solids in a turbulent protostellar disk with an application 
to the formation of icy planetesimals in the outer region 
of the nebula.

In this paper, I extend the previous studies presented
in HB03a$\&$b by considering the growth of a dust particle
due to interaction with the particulate background of a nebula
in the vicinity of its local density enhancement. The
main purpose of this paper is to study how the combined
effect of gas drag and pressure gradients can speed up
the coagulation of dust particles and the
formation of centimeter-sized objects. 

The model nebula considered in this paper is
presented in $\S$ 2. Section 3 presents the equations
of motion of a solid particle and  $\S$ 4 
deals with the analysis of the numerical results. 
Section 5 concludes this study by reviewing
the results and discussing their applications.

\section{The Model Nebula}

The model nebula in this study is
an isothermal and turbulence-free protostellar disk 
with a solar-type star at its center. The nebula is assumed 
to be a mixture of pure molecular hydrogen at hydrostatic
equilibrium,  and small
submicron-sized solid particles. It is also assumed that
the density of the gas maximizes at certain locations
in the nebula.

In a cylindrical
coordinate system with its origin at the location of the
central star and its polar plane on the midplane of the
nebula, the gas distribution function at any point $(r,z)$
is given by (HB03b)
\vskip 1pt
\begin{equation}
{\rho_g}(r,z)\,= \,{\rho_g}(r,0) \>{\rm {exp}}
\Biggl\{{{8GM}\over {\pi {{\bar v}_{\rm {th}}^2}}}\,
\biggl[{1\over {{({r^2}+{z^2})^{1/2}}}}-{1\over r}\biggr]\Biggr\}.
\end{equation}
\vskip 8pt
\noindent
In equation (1), $M$ is the mass of the central star,
$G$ is the gravitational constant, and 
${\rho_g}(r,0)$ represents the gas distribution function
on the midplane of the nebula.
For simplicity, the density of the gas
is considered to have an azimuthally symmetric maximum
on the midplane and is given by (HB03a$\&$b)
\begin{equation}
{\rho_g}(r,0)\,=\,{\rho_0}\,
{\rm {exp}}\Biggl[-\beta\Bigl(
{r \over{r_m}}-1 {\Bigr)^2}\Biggr]\,,
\end{equation}
where ${\rho_0}\,,{r_m}$ and $\beta$ are constant quantities
(Figure 1). 

The quantity ${{\bar v}_{\rm {th}}^2}=8{K_B}T/\pi {m_H}$ 
in equation (1) is the mean thermal velocity of the gas molecules.
In this equation, $K_B$ is the Boltzmann constant, 
$T$ is the temperature of the gas, and
$m_H$ is the molecular mass of hydrogen.
The gas density function
as given by equation (1) ensures that along the vertical
axis, the gravitational attraction of the central star will be
balanced by the vertical component of the pressure gradients
(HB03b). 

Because of their submicron sizes, the particles of the
background material of the nebula are
strongly coupled to the gas. The motions of these
particles are only affected by the gas drag. 
On the other hand, since
molecular hydrogen obeys the equation of state 
of an ideal gas, and because the nebula
is isothermal, any local enhancement in the gas density will
also enhance the pressure of the gas.
In the vicinity of such pressure enhanced regions,
any solid object undergoes radial migration toward the 
location of the maximum pressure while approaching
the midplane \citep[][HB03a$\&$b]{Whip64}. 
The rates of such migrations are small for small particles.
For the particles of the background, although
radial migration is extremely slow, 
along with vertical descent, it results
in accumulation of particulate material at the
locations of density enhancements and also in the
vicinity of the midplane. In this study, it is assumed
that at any position in the nebula, the gas and
particle distribution functions
are proportional. That is, 
${\rho_{\rm {dust}}}(r,z)=f\,{\rho_g}(r,z)$, where 
${\rho_{\rm {dust}}}(r,z)$ is the distribution
function of the background material. The
solid/gas ratio $f$ can attain different values
at different positions and can also vary with time. 
However, for simplicity, in this study, it is assumed to be 
a constant number much smaller than unity.

\section{Equation of Motion}

An object in the model nebula considered here is subject
to the gravitational attraction of the central star and
the drag force of the gas. Because of its small size, such an object,
similar to the particles of the background material, 
shows the tendency of staying with the gas 
and its dynamics is mostly driven by gas drag.
The tendency of solid particles in being coupled
to the gas is weaker for larger objects. 
For a particle with a size within the range
considered here, its
coupling to the gas is less strong than the
coupling of the submicron-sized particles of the 
background material. As a result, the particle
moves faster than the background material and
collides with them. Such collisions may result in
adhesion of the background particles to the moving object
and increase its mass. Such a change in the mass of the object
causes a change in its momentum and subsequently
affects its dynamics. 

The rate of the change of the momentum of an object
due to the sweeping of the background material is proportional
to the rate of the collision of the moving object with those particles. 
In general, whether a collision occurs 
between two objects depends on their relative velocity.
For an object with mass $m$ and radius $a$
at position $(r,z,\varphi)$ moving with velocity 
${\bf V}$ with respect to the central star, 
the rate of change of its momentum due to the sweeping
of the background particles is equal to
\begin{equation}
{{d{\bf P}}\over {dt}}=\pi\, {\rho_{\rm{dust}}}(r,z)\,
{a^2}\,{{d\ell}\over {dt}}\, ({{\bf V}-{\bf U}}),
\end{equation}
\noindent
where the sticking coefficient is 
equal to unity. In equation (3), ${\bf P}$
is the portion of the total momentum of the object that
changes by its coalescence with smaller background particles
and ${d\ell}=[{(dr)^2}+{(dz)^2}+{(rd\varphi)^2}]^{1/2}$ is
a line element along the path of the object.
The quantity ${\bf U}$ in equation (3) represents the
velocity of the particles of the medium along 
the line element $d\ell$.
In writing equation (3), it has been assumed that, 
while sweeping smaller particles, the
density of the object remains unchanged and it stays perfectly
spherical. It is important to emphasize
that in this equation $m$ and $a$ are functions of time,
and are related as
\begin{equation}
{{dm}\over {dt}}=\pi\, {\rho_{\rm{dust}}}(r,z)\,
{a^2}\,{{d\ell}\over {dt}}.
\end{equation}
\noindent
Equation (4) immediately implies
\begin{equation}
{{da}\over {dt}}={1\over 4}
\Bigl({{\rho_{\rm {dust}}}\over \rho}\Bigr)
{{d\ell}\over {dt}},
\end{equation}
\noindent 
where $\rho$ is the density of the object.

For small particles such as those
studied here, the magnitude of the velocity ${\bf V}$
is mainly dominated by its components in the radial
and vertical directions. As shown in HB03a$\&$b,
these components have larger values for larger objects.
That implies, while an object grows by sweeping the
particles of the background, it will move faster
in the radial and vertical directions. However,
due to the strong coupling to the gas,
for the particles of interest in this paper, 
the rates of increase in radial and vertical motions are quite small.
As a result, the magnitude of the velocity of the object
relative to the background material
(i.e., $|{\bf V}-{\bf U}|$), is also small.
This implies that, when migrating toward the
location of a maximum gas density, the
object approaches the background particles very slowly.
Such a slow approach suggests a gentle encounter between the
moving object and the particle of the background followed
by the sweeping up of the smaller particle by the larger object. 
Laboratory experiments have indicated that the collisional 
coagulation of millimeter-sized and smaller particles
can well be approximated by such a sweeping process
\citep{Wurm98}.

As mentioned above, the small velocities of the background particles
of the medium are the result of a strong coupling between the gas and
these particles. For these particles
such a coupling is so strong that with a
good approximation, one can replace ${\bf U}(r,z,\varphi)$ in 
equation (3) with the velocity of the gas
at that position. As a result, equation (3) can be
written as
\begin{equation}
{{d{\bf P}}\over {dt}}=\pi\, {\rho_{\rm{dust}}}(r,z)\,
{a^2}\,{{d\ell}\over {dt}}\, {\bf V}_{\rm {rel}},
\end{equation}
\noindent
where ${{\bf V}_{\rm {rel}}}$ 
is the relative velocity of the object with respect
to the gas. In the cylindrical coordinate system
considered here, the radial, vertical and tangential
components of this velocity are, respectively, given by 
${\dot r},\, {\dot z}$, and $r({\dot \varphi}\,-\,{\omega_g})$,
where the motions of gas molecules along the $z$-axis 
have been neglected. Because the gas is at hydrostatic
equilibrium, its angular velocity,
$\omega_g$, is slightly different from its Keplerian
value and is given by
\begin{equation}
{\omega_g^2} ={{GM}\over {({r^2}+{z^2})^{3/2}}} \,+\,
{1\over {r{\rho_g}(r,z)}}\,
{{\partial{{\cal P}_g}(r,z)}\over {\partial r}}.
\end{equation}
\vskip 5pt
\noindent
In this equation, ${{\cal P}_g}(r,z)$ is the pressure of the gas.

Considering the mass-growth equation (4), 
the equation of motion of an object in our model nebula
is given by
\begin{equation}
m{\ddot{\bf R}}\,=\,
-\,{{GMm}\over {({r^2}+{z^2})^{3/2}}}\,{\bf R}\,-\,
\pi\, {\rho_{\rm{dust}}}(r,z)\,
{a^2}\,{{d\ell}\over {dt}}\, {\bf V}_{\rm {rel}}\,-\,
{{\bf F}_{\rm {drag}}}\,,
\end{equation}
where ${\bf R}(r,z)$ is the position vector of the particle
and
\begin{equation}
{\bf F}_{\rm {drag}}= {4\over 3}\pi{{a^2}\over{a+\lambda}}
{{\bar v}_{\rm{th}}}
\Bigl[\lambda{\rho_g}(r,z)+{{3{m_0}}\over {2\sigma}}\Bigr]
{{\bf V}_{\rm {rel}}}\,,
\end{equation}
\noindent
represents the drag force of the gas. In equation (9),
$\sigma$ is the collisional cross section between two hydrogen
molecules and $\lambda$ is their mean free path. 

Equation (9) has been written following
\citet{Sup00} and HB03a$\&$b whose
authors present ${\bf F}_{\rm {drag}}$ in a form that
combines Epstein and Stokes drags in one formula. Because
the focus of this paper is on the growth of 1-1000
micron-sized objects to 1 centimeter in radius, 
as shown in HB03$\&$b, the gas 
Reynolds number (Re) for these objects in
the model nebula presented here stays well below
unity. This implies that the drag coefficient $C_D$ in 
${\bf F}_{\rm {drag}}$ has to be taken to be equal to 24/Re
\citep[][HB03a$\&$b]{Sup00,Weid77}.

\section{Numerical Results}

As mentioned in $\S$3, the distribution of the 
background material is considered to be proportional 
to the gas distribution function. \citet{Pod74} considered 
$f=0.0034$ for a solar nebula with fully formed silicates and
metals \citep[also see][]{Weid88}. \citet{Sup00},
however, have considered $f=0.0045$ by assuming that
the abundance of ${{\rm H}_2}{\rm O}$ in the nebula
is the same as its solar value. They call this value
of $f$ nominal and have also considered
$f=0.045$ and 0.45 in order to study the effect of an
increase in the concentration of material on the
midplane, on collision and coagulation of small solids.
The value of $f$ in this study is chosen to be
0.0034. From equation (4), it is evident that
larger values of $f$ will result in faster growth
of particles.

The equation of motion of a particle [Eq.(8)] and the
growth equation (5) were integrated, numerically,
for initial radii ranging from 
1 to 1000 microns, and for different values of the 
gas temperature. In all
these computations, the mass of the central star
was chosen to be equal to the mass of the Sun,
$\beta =1\,,{r_m}=1$ AU, and
${\rho_0}={10^{-10}}$g cm$^{-3}$. The collisional
cross section of hydrogen molecules, $\sigma$, was
taken to be $2\times {10^{-15}}\,{\rm {cm}}^{-2}$,
and their mean free paths 
$\lambda\,{\rm {(cm)}}=
4\times {10^{-9}}/{\rho_g}(r,z)\,({\rm {g cm}}^{-3})$.
The initial value of the height of a particle above the 
midplane was set equal to 1/10 of its initial
radial distance from the Sun.
The initial velocity of the particle along the
vertical axis was taken to be zero, and in the radial
direction, the particle was given an initial
Keplerian circular velocity. 

The growth of an object was followed until it reached
1 cm in size. Although centimeter-sized objects can also grow
larger by sweeping smaller particles, a 1 cm radius
was chosen as the upper limit since the assumed
growth process, that is, sweeping of smaller particles,
is more effective for dust grains and micron-sized objects.
In a nebula with centimeter-sized particles, in
addition to sweeping of the submicron particles of the background,
there is also collision among centimeter-sized particles
which may result in their coagulation and/or
fragmentation. Such collisions can also generate dust particles.
The collisions of micron-sized particles, on the other hand,
are gentle and result in coalescence of the involved objects.
To avoid the complexities associated with treating centimeter-sized
particles then, the numerical simulations were stopped at $a=1$ cm.
Figure 2 shows the growth of a 10 micron-sized grain
with a density of 2 g cm$^{-3}$, 
initially at $(r=3,z=0.3)$ AU. The gas temperature is 300 K.
As shown here, in a very short time, the particle
grows to 1 centimeter in size. During this time,
it undergoes radial migration
and also descends toward the midplane (Fig. 3).
For a comparison, the particle's
migrations without mass-growth have also been plotted.
As expected, the rates of the radial migration and
vertical descent of the particle increase by increasing its size.
Figures 3 and 4 clearly show that the combined effect of gas
drag and pressure gradients causes particles to rapidly
grow during their gas-drag induced migrations. 

A comparison between the two graphs of Figure 3 indicates that
during the time that the particle grows to 1 cm, it
descends toward the midplane faster than it migrates radially.
The actual path of the particle during this time has been shown
in Figure 4. Such a rapid vertical descent was also reported
by \citet{Nakagawa86}, where the authors have shown that,
for settling dust particles, the ratio of the vertical
and radial components of velocity is greater than one
and it increases by increasing the size of the particle.
The effect of
the radial component of the particle's velocity becomes
more pronounced when the particle reaches 1 cm in radius
and grows larger. \citet{Nakagawa86} call this phase
of the growth process the ``{\it{gas-dominated}}'' phase.
During this phase, the particle undergoes an over-damped
oscillatory motion in the vicinity of the midplane (HB03b).

To better understand the relation between the growth
of the particle and its vertical and radial motions,
it is useful to calculate the time and the distance
of collision between the object and a submicron grain 
of the background material. To define these quantities, recall that
it has been assumed that all collisions between
the object and particles of the background result in 100$\%$
sticking. The time of collision is therefore defined as the time
during which the mass of an object, $m(t)$, is increased
to $m(t)+{m_0}$, where $m_0$ is the mass of a background grain.
Because it has been assumed that after a collision,
the object maintains its spherical shape and will have
the same density, assuming a spherical shape for a 
background particle, and also assuming that it has
similar density as the moving object, after a collision
the radius of the object changes from $a(t)$ to
$[{a^3}(t)+{a_0^3}]^{1/3}$, where $a_0$ is the radius
of a background grain. The rate of change of radius, 
on the other hand, is given by equation (5). From this
equation and also considering that ${a_0}<<a(t)$,
the time of collision can be approximately written as
\begin{equation}
{\tau_c}(t)\,\simeq\,{{4{a_0^3}}\over {3{a^2}(t)}}\,
\Bigl({\rho\over {\rho_{\rm dust}}}\Bigr)\,
{\Bigl({{d\ell}\over {dt}}\Bigr)^{-1}}\,.
\end{equation}
\noindent
Equation (10) can also be obtained by introducing an
equivalent of a mean free path for the object. The
mean free path of the object, shown by $\lambda_d$, 
is defined as the distance that the object travels
before it sweeps a background grain. Similar to the
equation of the mean free path of a gas molecule, 
$\lambda_d$ can be written as
\begin{equation}
{\lambda_d}(t)\,=\,{{m_0}\over 
{\pi{[a(t)+{a_0}]^2}{\rho_{\rm dust}}}}\,.
\end{equation}
\noindent
Equation (11) immediately results in equation (10),
noting that ${\lambda_d}=(d\ell/dt){\tau_c}$.

Figure 5 shows the time of collision and the mean free path 
of the 10 micron object of Figure 2 during its growth to 1 cm. 
As shown here, the magnitudes of $\tau_c$ and $\lambda_d$ rapidly
decrease in time. To show that in more detail, the time of
growth has been divided into 4 segments.
As the object sweeps the background
material and grows in size, the rate of its gas-drag induced
migration becomes faster. As a result, it approaches a
background particle in a shorter time. Also,
as shown in Equation (11), the mean free path of
the object is inversely proportional to the square of 
its instantaneous radius. Since $a(t)$ increases
at all times, $\lambda_d$ is a monotonically
decreasing function of time.

Numerical simulations were also carried out for
different values of the gas temperature. Figure 6
shows the time of the growth of a 10 micron particle to 1 cm
for three different values of the temperature.
As shown here, the time of growth increases
at higher temperatures. That can be attributed to the 
fact that the spatial distribution of the background
material is directly proportional to the gas density
function [Eq. (1)]. By increasing the gas temperature, the
FWHM of $\rho_g$ and $\rho_{\rm dust}$ increases (Fig. 7)
indicating an increase in the spatial distances 
between the background particles. This causes the
time of collision to increase and the object will
take a longer time to grow larger at higher temperatures.

\section{Conclusions}

The results of the numerical study of the growth rates
of small particles in the vicinity of a local density
enhancement of a solar nebula were presented. It was shown
that, as a consequence of migration toward the
location of a local density enhancement,
solid objects can rapidly grow in size by sweeping up
the smaller particles of the background material.
Particles of interest in this study had sizes
between 1 to 1000 micron and their growth-rates were
studied until they reached 1 cm in radius. 
The results presented here indicate that, compared to
the time of settling and growth of dust particles
in a nebula without local gas density enhancements
\citep{Safronov,Weid80,Nakagawa81,Weid88}, 
in the vicinity of a local maximum gas density,
the combined effect of gas drag
and pressure gradients can increase the growth rates
of dust grains by one order of magnitude.

In this study the focus was on the growth of small grains to
objects of 1 cm in size. Although centimeter-sized objects  
can still acquire mass and grow larger
by sweeping up smaller particles, their dynamics is not
solely driven by gas drag and the sweeping process.
The velocities of such particles are large enough 
to cause them to fragment and produce dust 
when they collide. On the other hand, as noted by 
\citet{Safronov}, \citet{Weid80}, \citet{Nakagawa81},
\citet{Nakagawa86}, and \citet{Weid88},
the largest that dust particles can grow to while settling
toward the midplane is centimeter-sized.
The concentration
of centimeter-sized particles in the vicinity of the
midplane creates a turbulent layer \citep{Weid80} which
requires a more detailed analysis.
All these introduce complexities in treating centimeter-sized
objects that are unnecessary 
for the purpose of this study. Studies are, however, underway in which
the mutual interactions of centimeter-sized particles 
and the effect of turbulence on their collision and
coalescence have been considered.

The spatial density of the particulate background of the nebula was
chosen to be proportional to the gas distribution function.
The solid/gas ratio $f$ was chosen to be constant
and equal to 0.0034. This is a minimal value for $f$
that corresponds to a dust composition of metals and silicates
on the midplane of a standard nebula at 1 AU
\citep{Pod74,Weid80,Weid88}. In a more realistic
scenario, the value of $f$ is a function of 
position and time. For instance, in the model
nebula considered here, as time passes, the 
background material of the nebula settles toward
the midplane and also migrates toward $r=1$ AU,
where the gas distribution function maximizes.
This increases the dust/gas ratio around $r=1$ AU
on the midplane
and decreases $f$ at other locations. In such cases,
one has to take the time variation
of $f$ into consideration. However,
because in this study, the rates of radial and vertical 
migration of the background material are smaller than
those of the objects of interest, 
and also because it has been assumed that the
sticking probability is 100$\%$, which results in
faster growth of the particle, growing to
1 cm in size occurs so rapidly that 
the particles of the background 
will not have enough time to undergo appreciable
radial and vertical migrations. 

As mentioned earlier, the focus of this study has been
on the growth of particles to 1 cm in size.
For such small objects, the assumption of growth merely
due to sweeping up smaller particles is quite plausible
since these objects are coupled to the gas and they
approach one another slowly. Once the sizes of objects
increase 1 cm, their collisions may result in adhesion 
and/or fragmentation. As a result, their growth will not 
be entirely due to sweeping up of smaller particles and
will be affected by their individual velocities. 
Replacing the relative velocities of such objects
with their velocities relative to the gas is no longer 
a valid approximation. In such cases, one has to
follow individual particles and consider their 
velocities relative to one another.
Such an extension of this work
is currently in preparation for publication.

\acknowledgments

I am thankful to Alan Boss for critically reading the
original manuscript.
This work is partially supported by the NASA Origins of the Solar 
System Program under Grant NAG5-11569 and by the NASA Astrobiology
Institute under Cooperative Agreement NCC2-1056.

\clearpage

\begin{figure}
\plotone{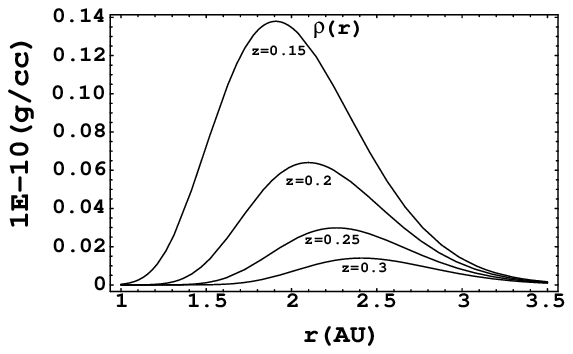}
\plotone{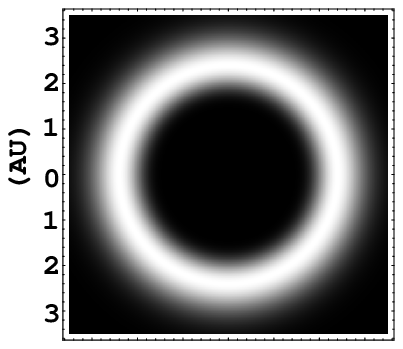}
\plotone{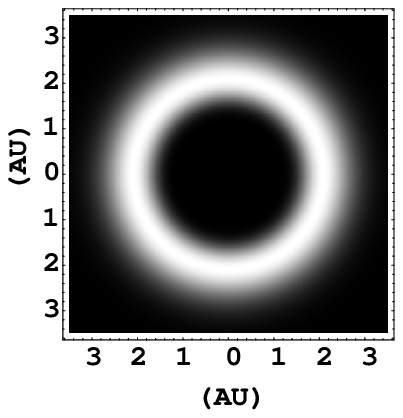}
\vskip -9.1in
\caption{Graphs of the gas distribution function.
The temperature of the gas is 300 K, $r_{\rm m}$=1 AU,
$\beta=1$, and ${\rho_0}={10^{-10}}$ g cm$^{-3}$.
The top graph depicts the radial distribution of the gas
at different heights above the midplane. The middle graph
shows a topview of the nebula at $z=0.3$ AU and the bottom
graph shows its topview at $z=0.15$ AU. 
\label{fig1}}
\end{figure}

\clearpage

\begin{figure}
\plotone{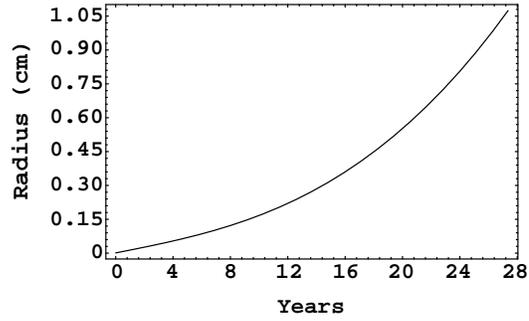}
\vskip -2.5in
\caption{Graph of the radius versus time 
for a particle with an initial
radius of 10 micron. The initial position 
of the particle is (3,0.3) AU, and the physical
properties of the gas are similar to those of Figure 1.
\label{fig2}}
\end{figure}

\clearpage

\begin{figure}
\plotone{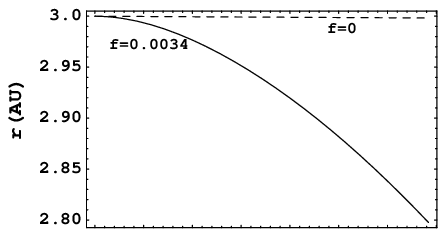}
\plotone{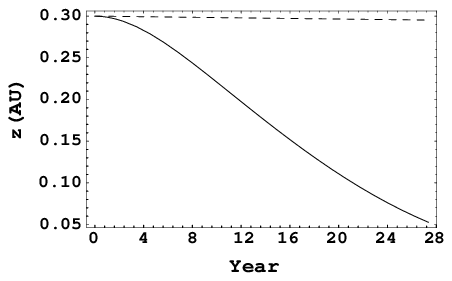}
\vskip -3in
\caption{Graphs of the radial migration (top) and 
vertical descent (bottom) of the 10 micron-sized particle 
of Figure 2. The solid lines
correspond to $f=0.0034$, and the dashed lines
represent the radial and vertical motions with no
mass-growth $(f=0)$. The horizontal axis
represents the time of the growth of the particle 
to 1 centimeter.
\label{fig3}}
\end{figure}

\clearpage

\begin{figure}
\plotone{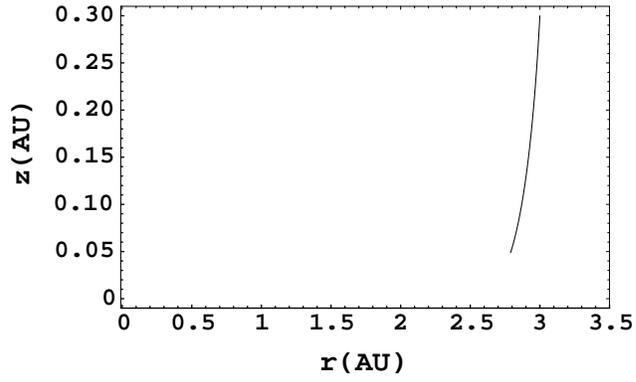}
\vskip -2in
\caption{The path of the 10 micron-sized particle
of Figure 3. As shown here, during its growth to
1 centimeter, the motion of the particle is mostly
vertical.
\label{fig4}}
\end{figure}

\clearpage

\begin{figure}
\plottwo{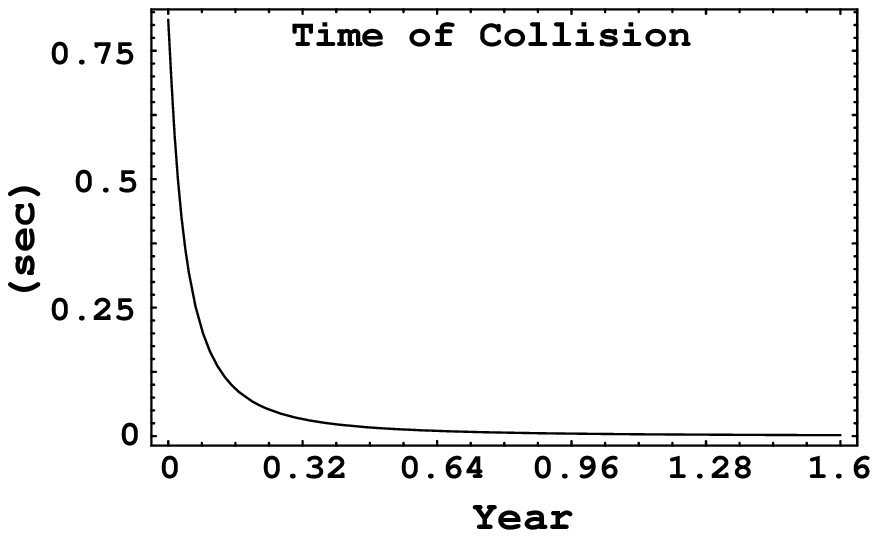}{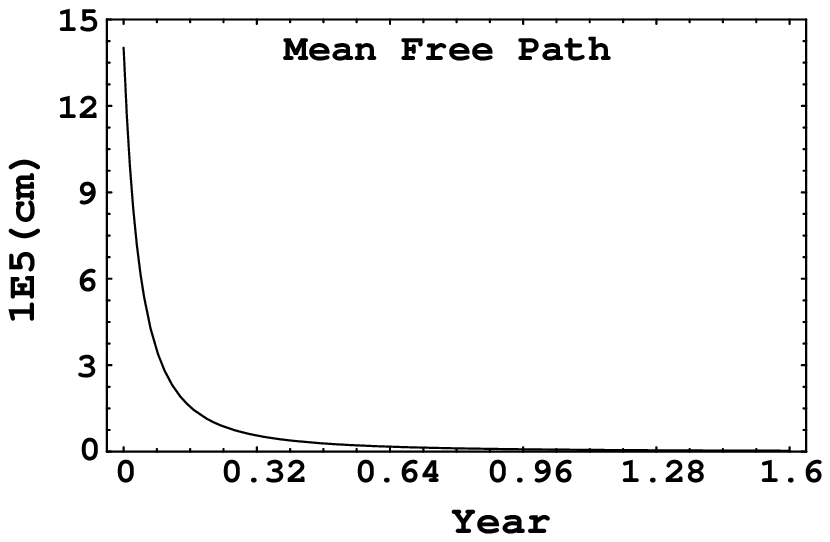}
\end{figure}
\begin{figure}
\plottwo{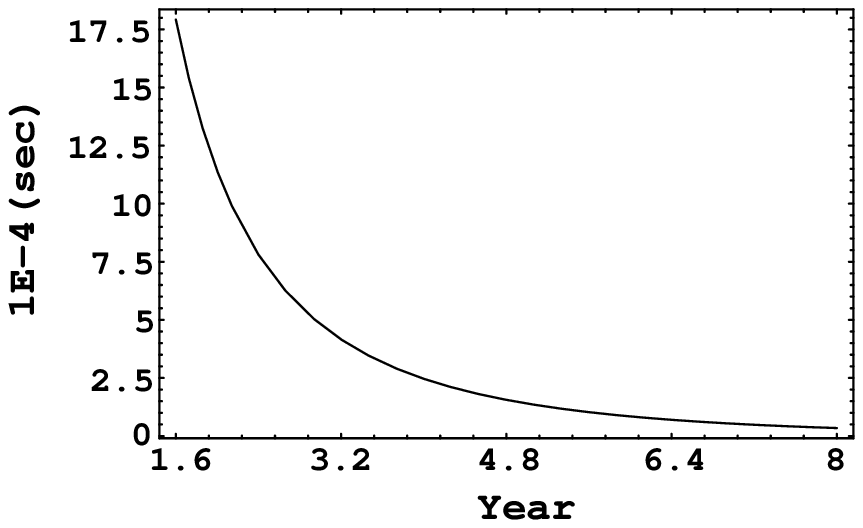}{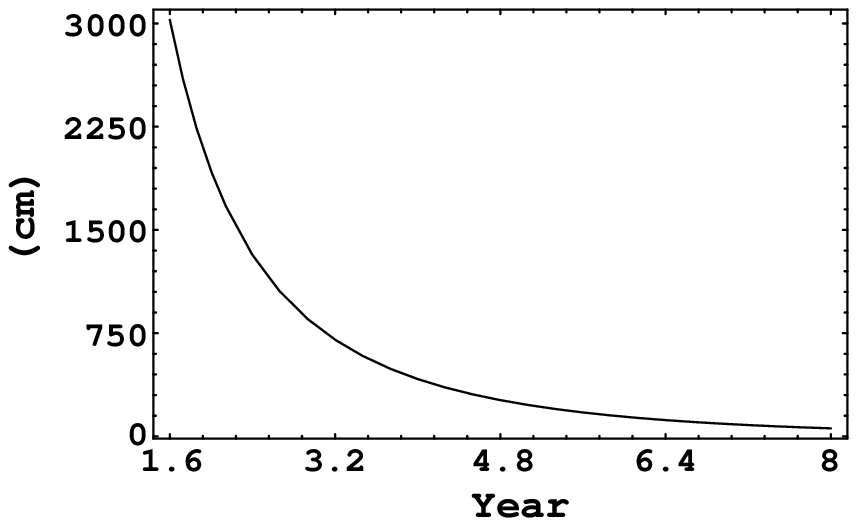}
\end{figure}
\begin{figure}
\plottwo{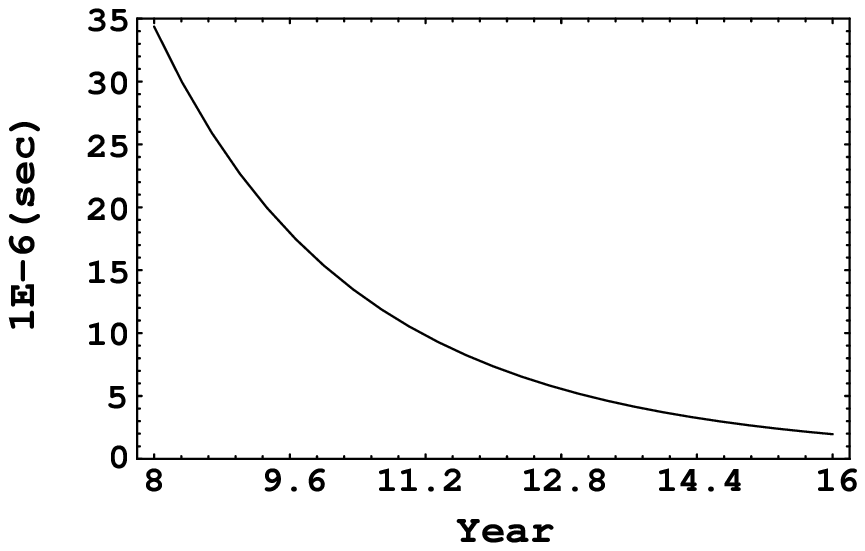}{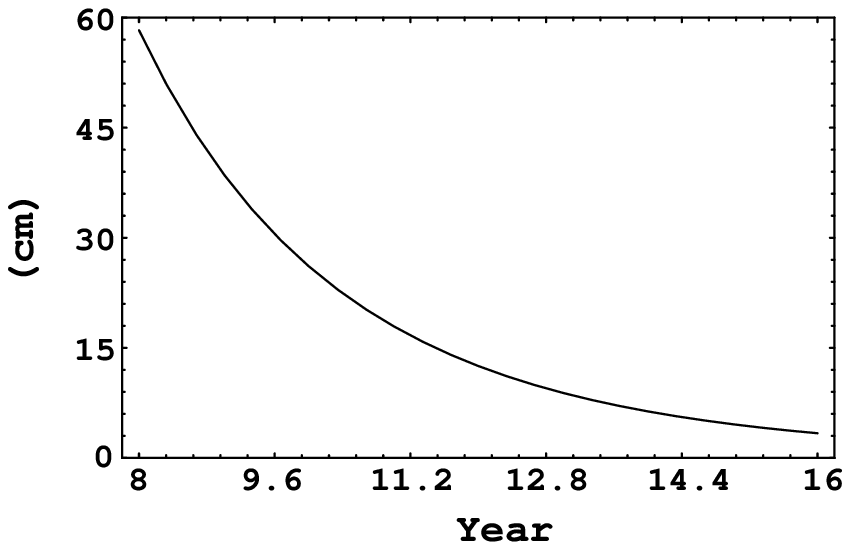}
\end{figure}
\begin{figure}
\plottwo{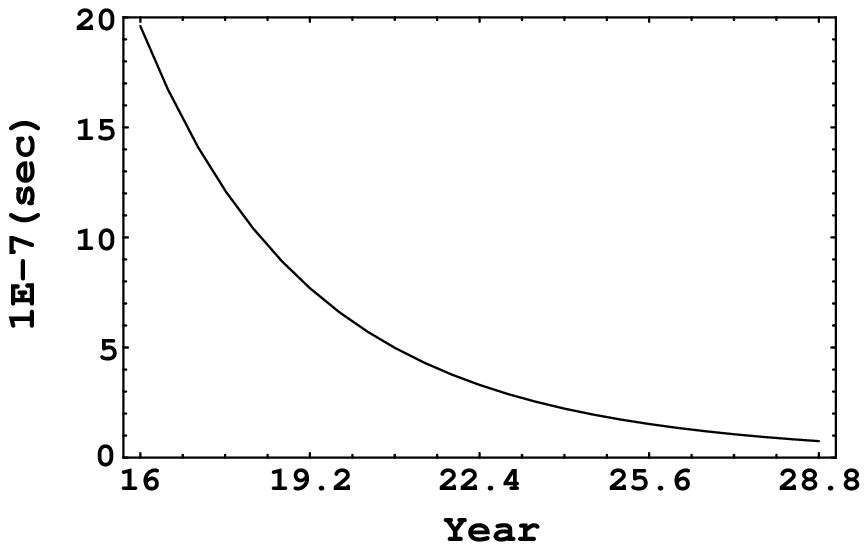}{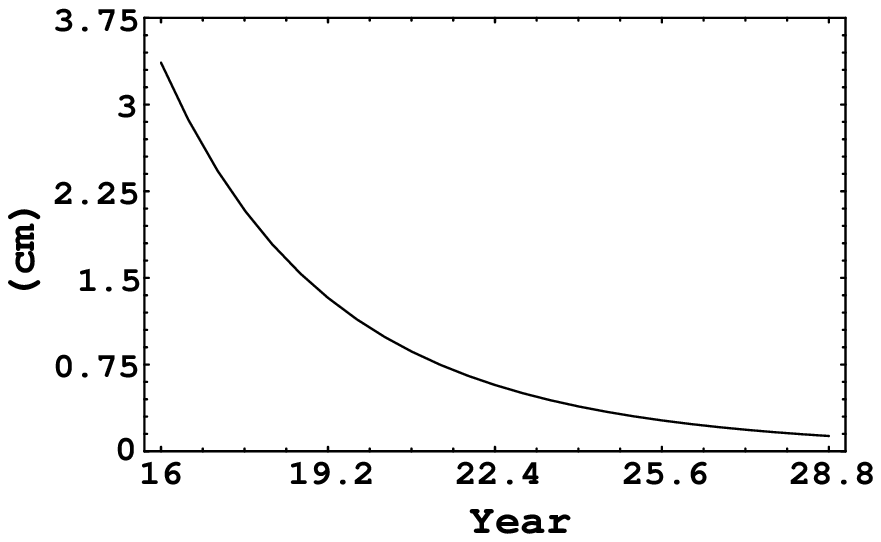}
\vskip -0.7in
\caption{Graphs of the time of collision between a growing
object and a particle of the background (left), and
the mean free path of the object (right). The horizontal
axes represent the time of growth. The initial radius 
of the object was 10 micron, and it was place at (3,0.3) AU.
The physical properties of the nebula are similar to those
of Figure 1. Note different scales on vertical axes.
\label{fig5}}
\end{figure}

\clearpage

\begin{figure}
\plotone{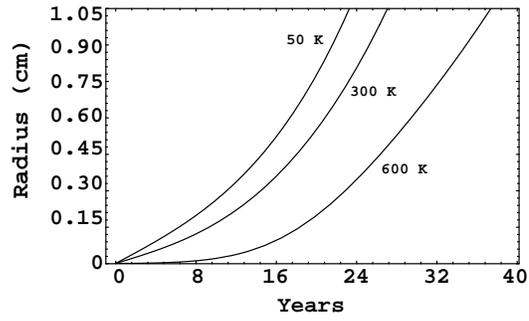}
\vskip -2.5in
\caption{Graph of the growth of a 10 micron-sized
particle with a density of 2 g cm$^{-3}$ 
for different values of the gas temperature. 
The particle was initially at (3,0.3) AU.
\label{fig6}}
\end{figure}

\clearpage

\begin{figure}
\plotone{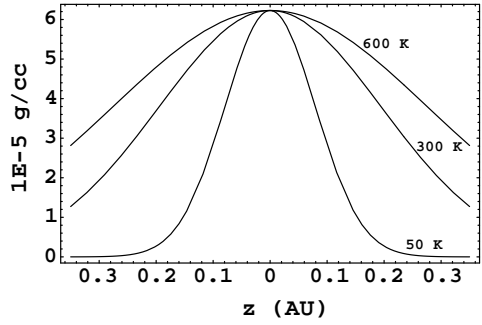}
\vskip -2.5in
\caption{Graph of the vertical distribution
of the background material at $r=3$ AU for different
values of the gas temperature.
\label{fig7}}
\end{figure}

\end{document}